\documentclass[10pt,sigconf,letterpaper,nonacm]{acmart}

\usepackage[labelfont={small,bf},textfont={small,it},skip=1pt]{caption}
\usepackage{graphicx}
\graphicspath{{figures/}{./}}
\makeatletter
\def\input@path{{sections/}{./}}
\makeatother
\usepackage{booktabs}
\usepackage{caption}
\usepackage{tabularx}
\usepackage[normalem]{ulem}
\usepackage{fancyref}
\usepackage{xcolor}
\usepackage{float}
\usepackage{pdflscape}
\usepackage{balance}
\usepackage{subcaption}

\useunder{\uline}{\ul}{}

\newcommand{\eg}{{\it e.g.}}

\title{Global Measurements of the Availability and Response Times of Public Encrypted DNS Resolvers}

\author{Ranya Sharma}
\affiliation{%
  \institution{University of Chicago}
  \country{}
}

\author{Nick Feamster}
\affiliation{%
  \institution{University of Chicago}
  \country{}
}

\begin{document}

\begin{sloppypar}

\begin{abstract}
Unencrypted DNS traffic between users and DNS resolvers can lead to privacy
and security concerns.  In response to these privacy risks, many browser
vendors have deployed DNS-over-HTTPS (DoH) to encrypt queries
between users and DNS resolvers.  
Today, many client-side
deployments of DoH, particularly in browsers, select between only a
few resolvers, despite the fact that many more encrypted DNS resolvers are
deployed in practice.
Unfortunately, 
if users only have a few choices of encrypted resolver, and only a few
perform well from any particular vantage point, then the privacy problems that
DoH was deployed to help address merely shift to a different set of third
parties. It is thus important to assess the performance characteristics of
more encrypted DNS resolvers, to determine how many options for encrypted DNS
resolvers users tend to have in practice.
In this paper, we explore the performance
of a large group of encrypted DNS resolvers supporting DoH by measuring DNS
query response times from global vantage points in North America, Europe, and
Asia.  Our results show that many non-mainstream
resolvers have higher response times than mainstream resolvers, particularly
for non-mainstream resolvers that are queried from more distant vantage
points---suggesting that most encrypted DNS resolvers are not replicated or
anycast.
In some cases, however, certain non-mainstream resolvers perform 
at least as well as mainstream resolvers, suggesting that
users may be able to use a broader set of encrypted DNS resolvers than those
that are available in current browser configurations.
\end{abstract}

\maketitle
\section{Introduction}\label{sec:intro}

The Domain Name System (DNS) is a critical component of the Internet's
infrastructure that translates human-readable domain names (\eg,
\texttt{google.com}) into Internet Protocol (IP) addresses~\cite{rfc1035}.
Most Internet communications begin with a client device sending DNS queries to
a recursive resolver, which in turn queries one or more name servers, which
ultimately refer the client to a server who can map the domain to an IP
address.
The \emph{response times} of these queries---the time 
to contact a recursive resolver, query various name servers, and return
the results---is important because the DNS underlies virtually all
communication on the Internet.  For example, loading a
web page, a browser must first resolve the domain names for each object on the
page before the objects themselves can be retrieved and rendered.  Thus, the
performance of DNS lookup is of utmost importance to application performance
such as web performance, as slow DNS lookup times will lead to slow web page
loads.

DNS did not originally take privacy and security into account: DNS queries
have historically been unencrypted, leaving users susceptible to
eavesdropping~\cite{schmitt2019:odns:pets}; queries can also be intercepted and
manipulated~\cite{jones2016detecting}.  To
address these types of privacy and security vulnerabilities, encrypted DNS
protocols have been developed and deployed, including DNS-over-HTTPS
(DoH)~\cite{rfc8484}, which is now deployed---and even enabled by default---in
many browsers.  DoH enables clients to communicate with recursive resolvers
over HTTPS, providing privacy and security guarantees that DNS previously
lacked.

For better or worse, most contemporary deployments of DoH have occurred in
browsers that provide limited options for
resolvers~\cite{ffChoices,chromeResolvers}.  Although DoH
protects against on-path eavesdropping, it does not prevent resolvers
\emph{themselves} from seeing the contents of DNS queries.  Thus, some have
argued that browser-based DoH deployments shift privacy concerns from
eavesdroppers to potential misuse by major DNS providers~\cite{vixie}.

\begin{table}[t]
\begin{footnotesize}
    \centering
    \addtolength{\tabcolsep}{-0.4em}
    \begin{tabular}{l|cccccc}
    \hline
    Browser & Cloudflare & Google & Quad9 & NextDNS & CleanBrowsing & OpenDNS
    \\
    \midrule
    Chrome    & \checkmark & \checkmark & & \checkmark & \checkmark & \checkmark \\
    Firefox  & \checkmark & & & \checkmark & & \\ 
    Edge   & \checkmark & \checkmark & \checkmark & \checkmark & \checkmark & \checkmark \\
    Opera            & \checkmark & \checkmark & & & & \\
    Brave            & \checkmark & \checkmark & \checkmark & \checkmark & \checkmark & \checkmark \\
    \bottomrule
    \end{tabular}
    \caption{Modern browsers provide only a few choices for encrypted DNS
    resolver, which we define as {\bf mainstream} resolvers.}
    \label{tab:SupportedResolvers}
\end{footnotesize}
\end{table}

Table~\ref{tab:SupportedResolvers} shows the DoH resolvers that have been deployed
to users of major browsers as of May 9,
2024~\cite{bravebrowser,edgebrowser,ffbrowser,chromebrowser,operabrowser}.  We
define the resolvers listed in Table~\ref{tab:SupportedResolvers} as {\em
mainstream}.
Yet, many other DoH resolvers have been deployed that are currently
not in use by major browser deployments~\cite{dnscrypt}---in other words,
there are many non-mainstream DoH resolver deployments.  

Previous studies have measured encrypted DNS performance, but they have mostly focused on mainstream DNS resolvers~\cite{hounsel2020comparing,hounsel2021can,hoang2020k,lu2019end-to-end}.
In this paper, we expand on these previous studies, exploring the performance
of all encrypted DNS resolvers---from a variety of global 
vantage points, as opposed to simply characterizing the mainstream DoH
providers from well-connected vantage points.
Towards this goal, we make the following contributions:
\begin{enumerate}
    \itemsep=-1pt
    \item We measure DoH response times a large list of resolvers, including
        both mainstream DoH resolvers that are included in major browser
        vendors {\em and} a large collection of non-mainstream resolvers.
    \item We study how the performance of various DoH resolvers differ based
        on vantage point.
    \item We study the performance of DoH resolvers in home networks. 
\end{enumerate}
\noindent
To our knowledge, this paper presents the first study of DoH performance
measurements for non-mainstream resolvers, as well as the first comparison of
DoH performance across a variety of vantage points, for a large number of
resolvers.
To perform these experiments, we developed and released an open-source
tool for measuring encrypted DNS performance to replicate and extend these
results, and to support further research on DoH performance.

The rest of the paper is organized as follows.  Section~\ref{sec:background}
provides background on DNS, including the origin of encrypted DNS and related
standards, and discusses related work.  Section~\ref{sec:method} details our research questions, the
experiments we conducted to study these questions, and the limitations of the
study.  Section~\ref{sec:results} presents the results of these experiments.
Section~\ref{sec:conclusion} concludes with a discussion of the implications
of these results and possible directions for future work.

\section{Background and Related Work}\label{sec:background}

In this section, we provide background on encrypted DNS protocols, including
the current deployment status of encrypted DNS, as well as various related
work on measuring encrypted DNS.

\subsection{Background: Encrypted DNS}

The Domain Name System (DNS) translates human-readable domain names into
Internet Protocol (IP) addresses, which are used to route
traffic~\cite{rfc1035}.  These queries have typically been unencrypted, which
enables on-path eavesdroppers to intercept queries and manipulate responses.

\paragraph{Encrypted DNS.}
Protocols for encrypting DNS traffic have been proposed, standardized, and
deployed in recent years, including DNS-over-HTTPS (DoH) and DNS-over-TLS
(DoT).  Zhu et al. proposed DoT in 2016 to address the eavesdropping and
tampering of DNS queries~\cite{zhu2015connection}.  It uses a dedicated port (853)
to communicate with resolvers over a TLS connection.  In contrast,
DoH---proposed by Hoffman et al. in 2018---establishes HTTPS sessions with
resolvers over port 443~\cite{rfc8484}.  This design decision enables
DoH traffic to use HTTPS as a transport, facilitating deployment as well as
making it difficult for network operators and eavesdroppers to intercept DNS
queries and responses~\cite{boettger2019empirical}. DoH can function in many
environments where DoT is easily blocked.

\paragraph{Moving the privacy threat.} Encrypting DNS queries and responses hides queries
from eavesdroppers but the recipient of the queries---the DNS resolver---can
see the queries~\cite{IEEEfight}. By design, recursive resolvers receive
queries from clients and typically need to perform additional queries to a
series of authoritative name servers to resolve domain names.  For these
resolvers to determine the additional queries they need to perform (or
determine if the query can be answered from cache), they must be able to see
the queries that they receive from clients.  Thus, although DoT and DoH make
it difficult for eavesdroppers along an intermediate network path to see DNS
traffic, recursive resolvers can still observe (and potentially, log) the
queries that they receive from clients.  The fact that many mainstream DoH
providers (e.g., Google) already collect significant information about users
potentially raises additional privacy concerns and makes it appealing for
users to have a large number of encrypted DNS resolvers that are reliable and
perform well. For this reason, users may wish to have more control over the
recursive resolver that they use to resolve encrypted DNS queries. Having
a reasonable set of choices that perform well in the first place is thus
important, and determining whether such a set exists is the focus of this
paper.

\paragraph{Status of browser DoH deployments.}
Most major browsers currently support DoH, including Brave, Chrome, Edge,
Firefox, Opera, and Vivaldi. Operating systems have also announced
plans to implement DoH, including iOS, MacOS, and
Windows~\cite{ffSettings,operaEdgeSettings,vivaldiSettings,iosSettings,jensen2020windows}.
In this paper, we focus on DoH because it is more widely deployed
than DoT.  Each of these browsers and operating systems either
currently support or have announced support for DoH (but, to our knowledge, not
DoT).

\subsection{Related Work}

\paragraph{Previous measurement studies of encrypted DNS.}
Previous studies have typically measured DoT and DoH response times the
protocols from the perspective of a few commonly used
resolvers~\cite{lu2019end-to-end}; in contrast, in this paper, we study a much
larger set of encrypted DNS resolvers, many of which are not available as
default options in major browsers.  Zhu et al. proposed DoT to encrypt DNS
traffic between clients and recursive resolvers~\cite{zhu2015connection}.
They modeled its performance and found that DoT's overhead can be largely
eliminated with connection re-use.  Böttger et al. measured the effect of DoT
and DoH on query response times and page load times from a university
network~\cite{boettger2019empirical}.  They find that DNS generally
outperforms DoT in response times, and DoT outperforms DoH.  They also find
that much of the performance cost for DoT and DoH can be amortized by re-using
TCP connections and TLS sessions.  

Hounsel et al. also measure response times
and page load times for DNS, DoT, and DoH using Amazon EC2
instances~\cite{hounsel2020comparing}.  They compare the recursive resolvers
for Cloudflare, Google, and Quad9 to the local recursive resolvers provided by
Amazon EC2 from five global vantage points in Ohio, California, Seoul, Sydney,
and Frankfurt.  They find that despite higher response times, page load times
for DoT and DoH can be \emph{faster} than DNS on lossy networks.  Lu et al.
utilized residential TCP SOCKS networks to measure response times from 166
countries and found that, in the median case with connection re-use, DoT and
DoH were slower than conventional DNS over TCP by 9 ms and 6 ms,
respectively~\cite{lu2019end-to-end}.
In contrast to previous work, our focus in this paper is not to measure the
DoH protocol itself or its relative performance to unencrypted DNS; instead,
our goal is to compare the performance of encrypted DNS resolvers {\em to
each other}, to understand the extent to which this larger set of DNS
resolvers could be used by clients and applications in different regions.

\paragraph{Studies and remedies for the centralization of encrypted DNS.} Other work has studied the centralization
of the DNS and proposed various techniques to address it.  Foremski et al.
find that the top 10\% of DNS recursive resolvers serve approximately 50\% of
DNS traffic~\cite{foremski2019dns-observatory}.  Moura et
al.~\cite{moura2020clouding} measured DNS requests to two country code
top-level domains (ccTLD) and found that five large cloud providers being
responsible for over 30\% of all queries for the ccTLDs of the Netherlands and
New Zealand.  Hoang et al.~\cite{hoang2020k} developed K-resolver,
which distributes queries over multiple DoH recursive resolvers, so that no
single resolver can build a complete profile of the user and each recursive
resolver only learns a subset of domains the user resolved.  Hounsel et al.
also evaluate the performance of various query distribution strategies and
study how these strategies affect the amount of queries seen by individual
resolvers~\cite{hounsel2021encryption}.
This line of research complements our work---these previous studies in many
ways motivate an enable the use of multiple encrypted DNS resolvers, but
designing a system to take advantage of multiple recursive resolvers must be
informed about how the choice of resolver affects performance.

\paragraph{Other DNS performance studies.} Researchers have also studied how
DNS performance affects application performance.  Sundaresan et al. used an
early FCC Measuring Broadband America (MBA) deployment of 4,200 home gateways
to identify performance bottlenecks for residential broadband
networks~\cite{sundaresan2013measuring}.  This study found that page load
times for users in home networks are significantly influenced by slow DNS
response times.  Allman studied conventional DNS performance from 100
residences in a neighborhood and found that only 3.6\% of connections were
blocked on DNS with lookup times greater than either 20 ms or 1\% of the
application's total transaction time~\cite{allman2020putting}. Otto et al.
found that the ability of a content delivery network to deliver fast page load
times to a client could be significantly hindered when clients choosing
recursive resolvers that are far away from CDN caches~\cite{otto2012content};
a subsequent proposal, namehelp, proxied DNS queries for CDN-hosted content
and sent them directly to authoritative servers.  Wang et al. developed WProf,
a profiling system that analyzes various factors that contribute to page load
times~\cite{wang2013demystifying}; this study demonstrated that queries for
uncached domain names at recursive resolvers can account for up to 13\% of the
critical path delay for page loads.  We do not measure the performance of a
large number of encrypted DNS resolvers from residential broadband access
networks in this paper nor do we study the effect of the choice of a broad
range of encrypted DNS resolvers on page load time, but doing so would be a
natural direction for future work.

\section{Method}\label{sec:method} In this section, we describe the metrics
used, how these metrics are measured, and our experiment setup.

\subsection{Metrics} 

\paragraph{Availability.} We are interested in determining which DoH
resolvers are still active and responding to queries.  We define a resolver as
unresponsive from a given vantage point if we fail to receive \emph{any}
response to the queries issued from a particular server.

\paragraph{Latency.} We
performed network latency measurements for each recursive resolver.  Each time
we issued a set of DoH queries to a resolver, we also issued a ICMP ping
message and noted the round-trip time.  This enabled us to explore
whether there was a consistent relationship between high query response times
and network latency.

\paragraph{DNS query response time.} We define DNS query response time as the
end-to-end time it takes for a client to initiate a query and receive a
response.  To measure query response times with various DoH resolvers, we
performd dig queries to the resolvers. We used the public, open-source Netrics
platform~\cite{netrics} to conduct periodic measurements. Netrics
provides a variety of open-source network measurement tests, and we added our 
test to this suite. Our tool enables researchers to issue
traditional DNS, DoT, and DoH queries.

Our tool supports continuous DoH response time measurements. Clients to provide
a list of DoH resolvers they wish to perform measurements with. After a set of 
measurements complete with a list of DoH resolvers and domain names, the tool
writes the results to a JSON file.

\subsection{Experiment Setup}
To provide a comparative assessment of DNS
performance across DoH resolvers, we perform measurements across 91 DoH
resolvers, grouped by their geographical locations—18 in North America, 13 in
Asia, and 33 in Europe~\cite{dnscrypt}. 6 resolvers were unable to return a location. 
We employed MaxMind's GeoLite2 databases to geolocate each DoH resolver~\cite{maxmind}.

We collect measurements via two sources—-home network devices and Amazon EC2 instances. 
Both sources use the same measurement platform. This allows us to explore measurements from a wide variety
of access networks. We performed continuous measurements on four devices deployed in a single apartment complex located in a 
Chicago metropolitan area between June 22--September 30, 2023. We designed our tests to run every few hours. They were measured over IPv4. We also performed our measurements on Amazon EC2 instances between September 19--October 16, 2023. 
After this 1 month experiment span, we continued to take measurements for 1-3 days per month until May 2024. 
The motivation behind this was to ensure that resolver performance did not change drastically since October 2023. 
Those measurement spans were February 8--February 10, 2024, 
March 12--March 13, 2024, April 12--April 14, 2024. Measurements were performed three times a day. 
We also took the four highest performing resolvers (\texttt{Google}, \texttt{Cloudflare}, \texttt{Quad9},
\texttt{Hurricane Electric}) located in North America and measured their
performance in Europe and Asia to better understand how they compare in
farther vantage points. 

\paragraph{Vantage Points.} 
We performed our measurements across four units in the same apartment complex in the Chicagoland area. 
We installed our measurement platform on Raspberry Pi devices, which we used to 
collect data. 

We also performed our measurements from three global
vantage points through Amazon EC2~\cite{amazon_ec2}.  We deployed one server
in each of the Ohio, Frankfurt, and Seoul EC2 regions.  We chose to perform
measurements from multiple global vantage points to understand how DoH
performance varies not only by which resolver is used, but also which
geographic region the client is located in.  Each server utilized 8 GB of RAM
and 2 virtual CPU cores (the \texttt{t2.xlarge} instance type), and they each
used Linux/UNIX~\cite{amazon_ec2_instance_types}.

\paragraph{Resolvers and Domain Names.} \Fref{sec:resolvers} lists each of 
the DoH resolvers we measured.  These resolvers were
scraped from a list of public DoH resolvers provided by the \texttt{DNSCrypt}
protocol developers~\cite{dnscrypt-public-resolvers}.
We chose this list of resolvers because it was publicly available.
Previous work has largely studied major DNS providers in use by web browsers; in contrast, we
measure the performance of a larger set of encrypted DNS resolvers~\cite{hounsel2020comparing,hounsel2021can,hoang2020k,lu2019end-to-end}.

We issued queries for three domains to each resolver: \texttt{google.com}, 
\texttt{amazon.com}, and \texttt{wikipedia.com}.  We chose these domains based on their popularity, but
other domain names would have likely sufficed.  We do not expect our choice of
domain names to unfairly skew our performance comparisons between resolvers. 

\paragraph{Caching}
Note that our selection of domains is quite realistic, because it is reasonable to expect that most people query sites that are already in cache. 
Therefore, in our method, we capture the typical user experience.
We are primarily interested in the behavior of a resolver and the presence of cached entries enables a more controlled experiment.

\paragraph{Measurement Procedure.} We performed the following steps to measure
the performance of each of the encrypted DNS resolvers, from each of our three vantage points:
\begin{enumerate} 
        \item For each resolver that we aim to measure, perform a dig query, measuring the query response time for three
            domain names.
    \item For each resolver, issue a ICMP ping
            probe and collect the round-trip latency. 
\end{enumerate}

\paragraph{Limitations.} Our work has several limitations.
First, we do not measure how encrypted DNS affects application
performance, such as web page load time. Ultimately, an assessment of the
effects of encrypted DNS performance on application performance, including web
page load time, across the full set of encrypted DNS resolvers, could provide
a more comprehensive understanding of the effects of encrypted DNS on
application performance. Furthermore, although we do not expect it to affect conclusions, it may
be informative to perform measurements from a larger set domain names; our
measurements perform DNS lookups to just three domain names.

\section{Results}\label{sec:results} 

In this section, we describe the results of our experiments.  We explore which
non-mainstream resolvers are available; how
the performance of mainstream resolvers compares to that of non-mainstream
resolvers; and whether (and to what extent) encrypted DNS response time
correlates to high network latency. We also explore how resolvers perform 
differently from home networks versus EC2 instances. 

\paragraph{Are Non-Mainstream Resolvers Available?}
We first aimed to study the availability of encrypted DNS
resolvers. 
We received responses from most resolvers that we queried. When trying to perform DoH queries from all vantage points,
we received 5,098,281 successful responses and 311,351 errors. 
The most common errors we received from attempting to communicate with
the resolvers were related to a failure to establish a connection.
We did not identify a consistent pattern of not receiving responses from a certain
subset of resolvers each time the measurements ran.

\begin{figure}[t!]
\centering\includegraphics[width=0.8\columnwidth]{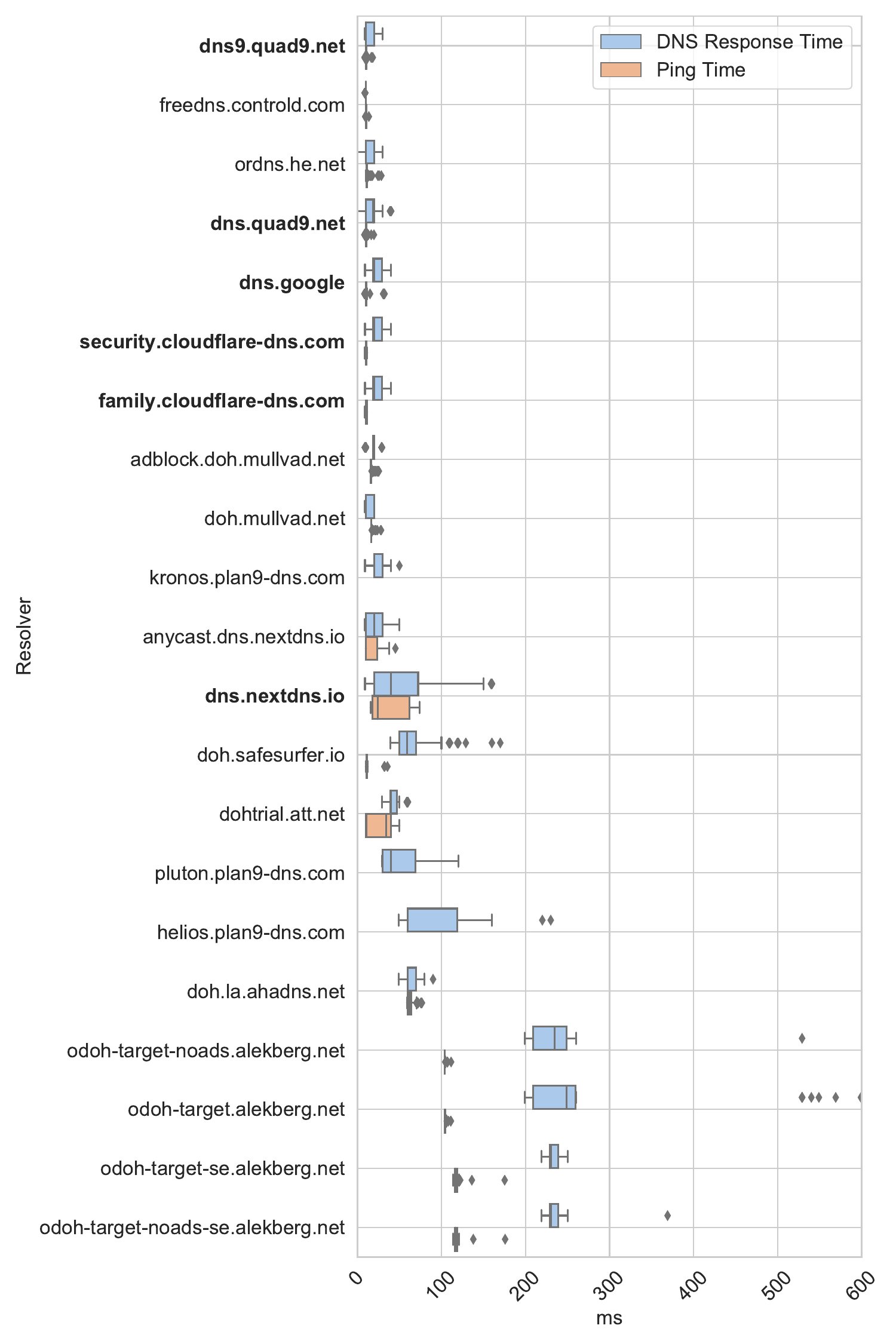}
\caption{The DNS response time and ICMP ping time distributions for
    encrypted DNS resolvers located in North America, measured from an EC2
    instance in Ohio.  The plot shows distributions for both DNS response times
    and ICMP round-trip latency. Mainstream resolvers are shown in boldface.
    Results for other vantage points are shown in Figures~\ref{fig:dns-NA}--\ref{fig:dns-europe} in the Appendix.}
    \label{fig:dns-us-ohio}
\end{figure}

\paragraph{How Do Non-Mainstream Resolvers Perform?}
Given the large number of non-mainstream resolvers that have not been
previously studied, we aimed to study how
the performance of these encrypted DNS resolvers compared to mainstream ones.
As previously mentioned, one of our motivations in doing so is to better
understand the global extent of encrypted DNS resolver deployment, as existing
lists of public encrypted DNS resolvers~\cite{dnscrypt-public-resolvers} do
not provide any overview of either reliability or performance. Additionally,
given that most organizations who have deployed mainstream resolvers are based
in the United States, we also wanted to explore how the performance of a
broader set of encrypted DNS resolvers varied by geography.

Figure~\ref{fig:dns-us-ohio} shows the distributions of DNS response times and
ICMP ping times across encrypted DNS resolvers located in North America, as
measured from an EC2 instance in Ohio. Note that some of these resolvers are anycasted,
so they are not exclusively located in North America. However, the purpose of this analysis
is more to understand how the resolvers perform from the perspective of the client.
The plots show distributions for both DNS response times and ICMP round-trip latency.  Although some distributions
extend beyond 600~ms, we have truncated the plots for ease of exposition,
since responses beyond this range will not result in good application
performance.  Certain resolvers did not respond to our ICMP ping probes; for
those resolvers, no latency data is shown.
As expected, most mainstream resolvers outperformed non-mainstream resolvers
from most vantage points.  Non-mainstream resolvers also exhibited higher
variability of median query response times.  
From the Chicago home network devices, response times from resolvers were as
high as 399~ms, identified as much above typical DNS lookup response times.
From the Ohio vantage point, the maximum response time from a resolver was 270~ms. 

We also explored the distributions of DNS response times and ICMP ping times
across encrypted DNS resolvers located across continents, as measured from
vantage points in the United States, Asia, and Europe, respectively.  From
Seoul, the maximum response time from a resolver was 569~ms; from Frankfurt,
the maximum response time from a resolver was 380~ms.  We observe more
consistent performance for resolvers located in Europe, but we still note high
variability from the Seoul vantage point.
Figures~\ref{fig:dns-NA}--\ref{fig:dns-europe} in the Appendix show all of
these results in more detail.

In some cases, one particular non-mainstream resolver would outperform
the mainstream DoH resolvers.  As expected, \texttt{dns.quad9.net},
\texttt{dns.google}, and \texttt{dns.cloudflare.com} were among the top five
highest performing DoH resolvers in North America, Europe, and Asia.
Interestingly, however, \texttt{ordns.he.net}---a DoH resolver hosted by
Hurricane Electric, a global Internet service provider (ISP)---managed to
outperform all mainstream resolvers from the home network devices. From Frankfurt, \texttt{dns.brahma.world}
outperforms \texttt{dns.cloudflare.com}; from Seoul, \texttt{dns.alidns.com} outperforms \texttt{dns.quad9.net},
\texttt{dns.google}, and \texttt{dns.cloudflare.com}; and from Ohio, \texttt{freedns.controld.com} outperforms \texttt{dns.google}
and \texttt{dns.cloudflare.com}.

We observe that the measurements performed from home network devices 
demonstrate fewer variability than the measurements performed from EC2 instances, 
especially the Seoul instance. Despite a few outliers, such as \texttt{doh.la.ahadns.net}, 
the graphs do not demonstrate high variability. In fact, \texttt{doh.la.ahadns.net} has significant response times and variability in the home network measurements, but very little variability in the EC2 measurements. It is worth noting that resolver performance can vary across measurements collected on virtual instances versus home networks. Furthermore, \texttt{dns.twnic.tw} demonstrates high ping times and response times from the home network measurements, but low times and variability from the EC2 measurements. Conversely, \texttt{antivirus.bebasid.com} has high variability from the Ohio and Frankfurt EC2 instances, but low variability from the home network devices. 
Except for these cases, the median resolver response times are almost identical for the home network and Ohio EC2 measurements. 

\begin{table}[t!]
\centering
\begin{tabular}{l|rr}
\toprule
    \textbf{Resolver} & \multicolumn{2}{c}{\textbf{Vantage Point}} \\
                  & \textrm{Seoul (ms)}         & \textrm{Frankfurt (ms)} \\
\midrule
antivirus.bebasid.com                                & 99 & 380                            \\
dns.twnic.tw                          & 59                                          & 290                              \\
dnslow.me                                & 29                                           & 240                              \\
jp.tiar.app                            & 39                                           & 250                             \\
public.dns.iij.jp                              & 39.5                                           & 250                               \\
\bottomrule
\end{tabular}
    \caption{Median DNS response times for non-mainstream resolvers (Asia).}
\label{tab:UnconvAsia}
\end{table}

\begin{table}[t!]
\centering
\begin{tabular}{l|rr}
\toprule
\textbf{Resolver} & \multicolumn{2}{c}{\textbf{Vantage Point}} \\
                  & \textrm{Frankfurt (ms)}     & \textrm{Seoul (ms)} \\
\midrule
doh.ffmuc.net                               & 70 & 569                         \\
dns0.eu                        & 20 & 399                         \\
open.dns0.eu         & 10 & 324                         \\
kids.dns0.eu                                 & 10 & 309                         \\
dns.njal.la                        & 20 & 289                         \\
\bottomrule
\end{tabular}
    \caption{Median DNS response times for non-mainstream resolvers (Europe).}
\label{tab:UnconvEur}
\end{table}

To better understand the extent to which certain encrypted DNS resolvers
perform well for clients in some regions but not others, we identified
resolvers that exhibited low DNS response times in for clients in one region
but not another. 
Tables~\ref{tab:UnconvAsia} and~\ref{tab:UnconvEur} show 
the five encrypted DNS resolvers for Europe and Asia that exhibit the 
largest differences in median DNS response times when
queried from a remote vantage point (queries of resolvers in Asia from Europe,
and of resolvers of Europe from Asia, respectively). In both cases, 
Table~\ref{tab:UnconvAsia} shows that non-mainstream resolvers located
in Asia perform better from the vantage point in Seoul than the one in
Frankfurt.  Similarly, as expected, Table~\ref{tab:UnconvEur} shows that the
median response times of non-mainstream resolvers in Europe are much lower when
measured from Frankfurt than the response times of those same resolvers
measured from Seoul.

\section{Discussion}\label{sec:discussion} 

These findings highlight promising advancements and areas for further development. 
Mainstream DoH deployments deployments by major browser vendors have evidently established a robust encrypted DNS infrastructure, with strong performance,
offering improved privacy and security.  
However, the broader ecosystem of non-mainstream encrypted DNS resolvers presents a more nuanced picture. 
Although these alternative resolvers offer greater diversity, they oftentimes suffer from slower response times, except for a few cases, particularly when they are not locally based. 
This underscores the challenges associated with ensuring global availability and consistent quality across the encrypted DNS landscape, while ensuring that queries are distributed across multiple encrypted resolvers.
Note that the presence of high-performing non-mainstream resolvers is promising and encouraging. These show that lesser-known providers can achieve similar reliability to mainstream resolvers, offering
users alternative choices. 

Overall, the encrypted DNS ecosystem should be expanded to include more globally distributed, high-performing alternatives that are operated by a broader range of organizations. 
This diversification is necessary to enhance user privacy and avoid the over-reliance on a handful of popular vendors. 
\label{lastpage}\section{Conclusion}\label{sec:conclusion}

The increased deployment of encrypted DNS, including DNS-over-HTTPS~(DoH) has
been accompanied both with ``mainstream'' DoH deployments in major browser
vendors, as well as a much broader deployment of encrypted DNS servers around
the world that are not among the common choices for resolvers in major
browsers.  Understanding the viability of this larger set of encrypted DNS
resolvers is important, particularly given that a lack of diversity of viable
resolvers potentially could create new privacy concerns, if only a small
number of organizations provided good performance. 
We find that many non-mainstream resolvers have higher median response times
than mainstream ones, particularly if the resolvers are not local to the
region; in contrast, most mainstream resolvers appear to be replicated and
provide better response times across different geographic regions. In some
cases, however, a local non-mainstream resolver can exhibit equivalent
performance as compared mainstream resolvers (\eg, {\tt ordns.he.net}, {\tt
freedns.controld.com},{\tt dns.brahma.world}, and {\tt dns.alidns.com}). These
results suggest both good news and room for improvement in the future: On the
one hand, viable alternatives to mainstream encrypted DNS resolvers do exist.
On the other hand, users need easy ways of finding and selecting these
alternatives, whose availability and performance may be more variable over
time than mainstream resolvers. It is also clear that there is an opportunity
to invest in deploying and maintaining reliable, performant, global encrypted
DNS infrastructure operated by a greater diversity of organizations.

\pagebreak
\bibliographystyle{abbrv}
\bibliography{paper}

\pagebreak
\appendix
\section{Appendix}

\subsection{Ethics}
This work does not raise any ethical issues and 
is determined to be non-human subjects research by our university's
Institutional Review Board.  The model of deployment is very much akin to a
``RIPE Atlas'' model, a longstanding mode for deploying measurement probes in
access networks.

\subsection{Resolvers}\label{sec:resolvers}
\begin{small}
\begin{itemize}
\item anycast.dns.nextdns.io
\item unicast.uncensoreddns.org
\item doh.ffmuc.net
\item jp.tiar.app
\item dns.therifleman.name
\item doh.pub
\item dns10.quad9.net
\item dns.adguard.com
\item doh.mullvad.net
\item dns12.quad9.net
\item dns-unfiltered.adguard.com
\item dns.alidns.com
\item helios.plan9-dns.com
\item dns1.ryan-palmer.com
\item dns.digitale-gesellschaft.ch
\item chewbacca.meganerd.nl
\item ordns.he.net
\item dns11.quad9.net
\item anycast.uncensoreddns.org
\item doh.libredns.gr
\item dns.brahma.world
\item dns.switch.ch
\item dns-doh-no-safe-search.dnsforfamily.com
\item ibksturm.synology.me
\item kronos.plan9-dns.com
\item dns-family.adguard.com
\item freedns.controld.com
\item dnsforge.de
\item dns-doh.dnsforfamily.com
\item public.dns.iij.jp
\item family.cloudflare-dns.com
\item dns.google
\item v.dnscrypt.uk
\item doh.dnscrypt.uk
\item doh.safesurfer.io
\item doh.la.ahadns.net
\item doh.tiar.app
\item doh.sb
\item doh-2.seby.io
\item dns.twnic.tw
\item dns.njal.la
\item pluton.plan9-dns.com
\item doh.seby.io
\item dns.quad9.net
\item dns.digitalsize.net
\item dns9.quad9.net
\item dohtrial.att.net
\item doh.nl.ahadns.net
\item adblock.doh.mullvad.net
\item adl.adfilter.net
\item per.adfilter.net
\item syd.adfilter.net
\item dns.nextdns.io
\item dns0.eu
\item doh.360.cn
\item open.dns0.eu
\item dnslow.me
\item kids.dns0.eu
\item pdns.itxe.net
\item security.cloudflare-dns.com
\item sby-doh.limotelu.org
\item dns.bebasid.com
\item 1dot1dot1dot1.cloudflare-dns.com
\item antivirus.bebasid.com
\item odoh-target-noads.alekberg.net
\item odoh-target-se.alekberg.net
\item odoh-target-noads-se.alekberg.net
\item odoh-target.alekberg.net
\item dnsse-noads.alekberg.net
\item dnsse.alekberg.net
\item family.puredns.org
\item dnsnl.alekberg.net
\item dnsnl-noads.alekberg.net
\item puredns.org
\item dns.circl.lu
\end{itemize}
\end{small}

\subsection{Response Time Measurements}

\begin{figure*}[h!]
\centering
\begin{subfigure}[b]{0.35\textwidth}
\includegraphics[width=\textwidth]{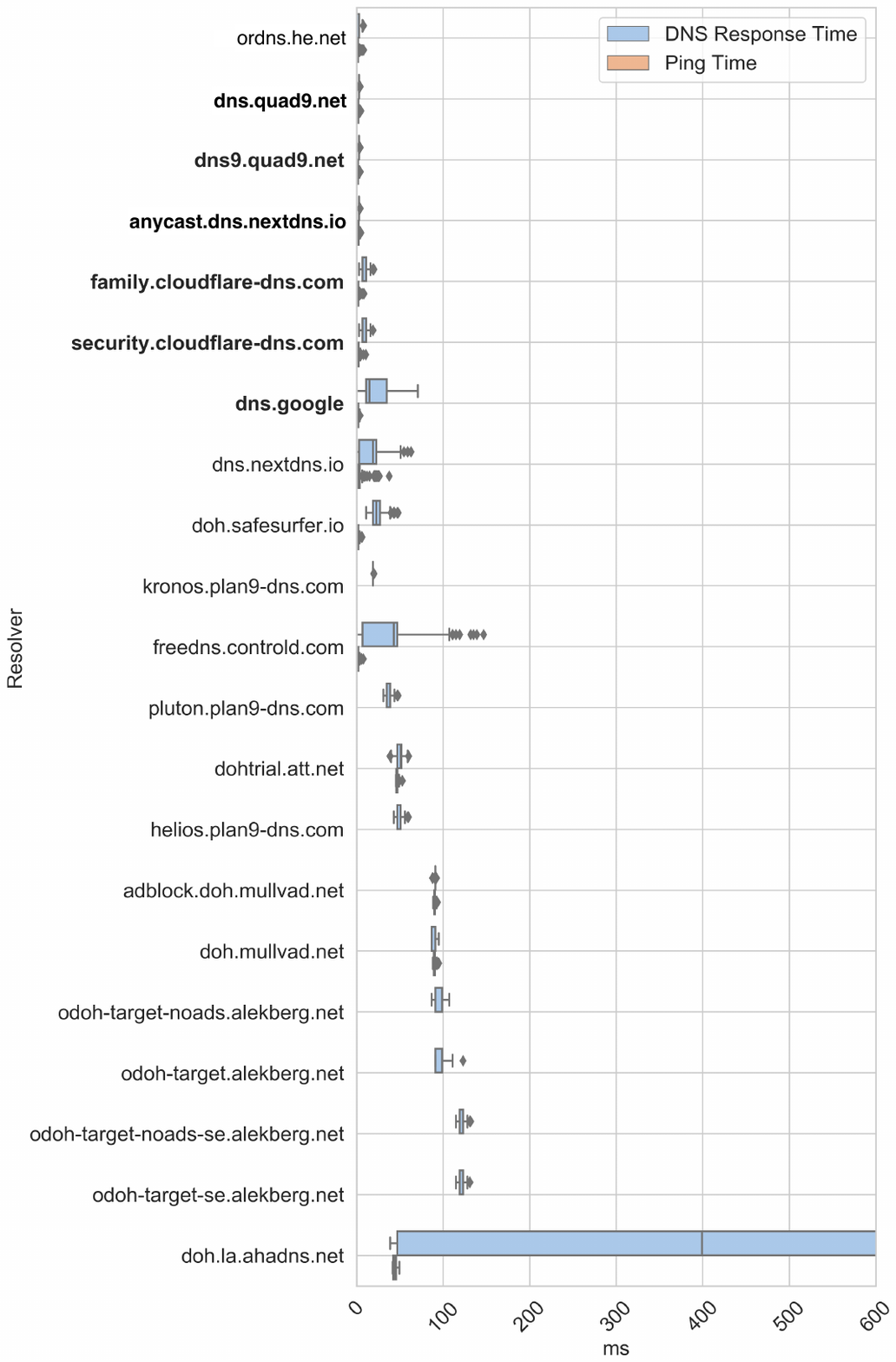}
    \caption{U.S. Home Networks (Local)}
\end{subfigure}
\begin{subfigure}[b]{0.35\textwidth}
\includegraphics[width=\textwidth]{ohio_NA.pdf}
    \caption{Ohio EC2 (Local).}
\end{subfigure}
\hfill \\
\begin{subfigure}[b]{0.35\textwidth}
\includegraphics[width=\textwidth]{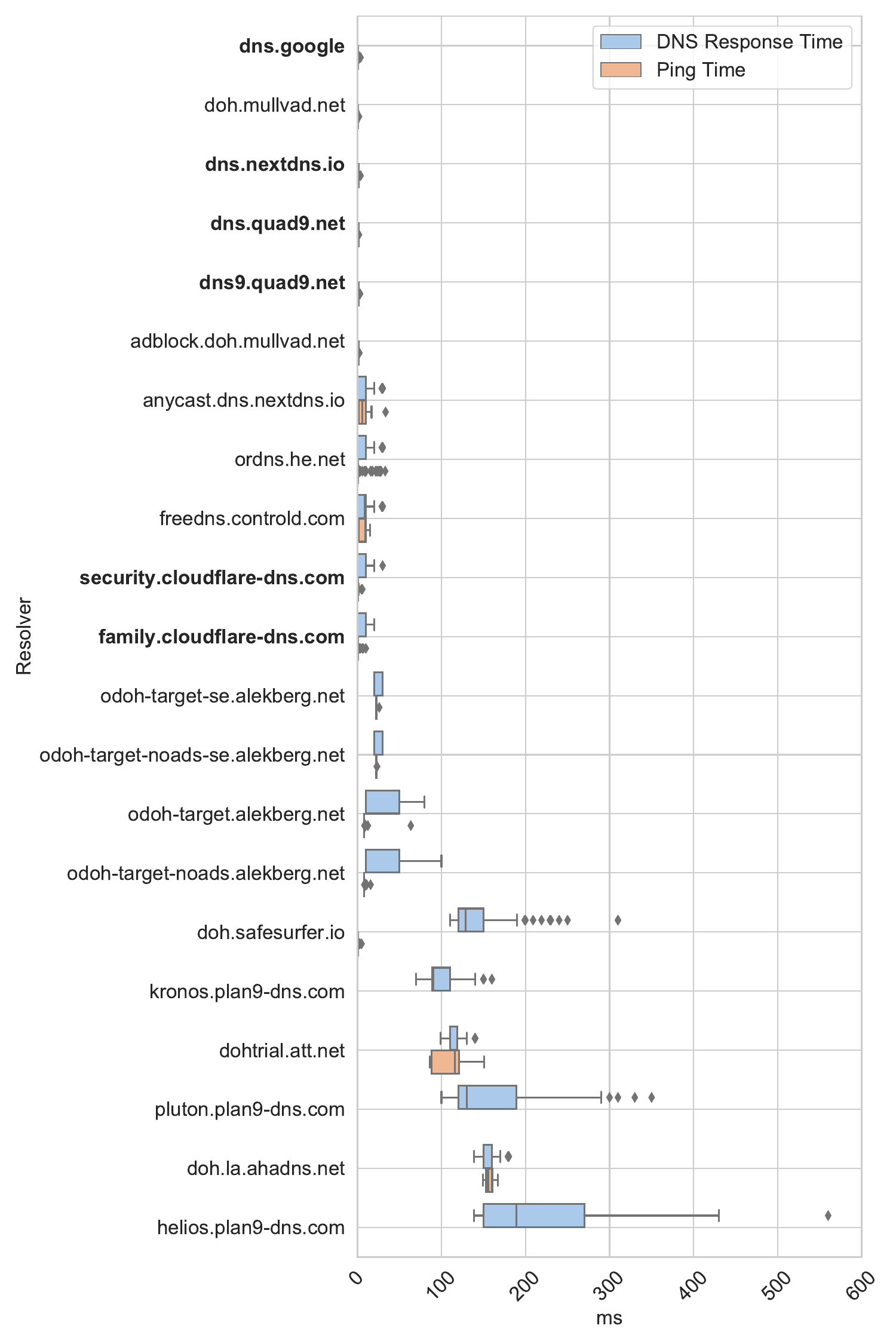}
    \caption{Frankfurt EC2.}
\end{subfigure}
\begin{subfigure}[b]{0.35\textwidth}
\includegraphics[width=\textwidth]{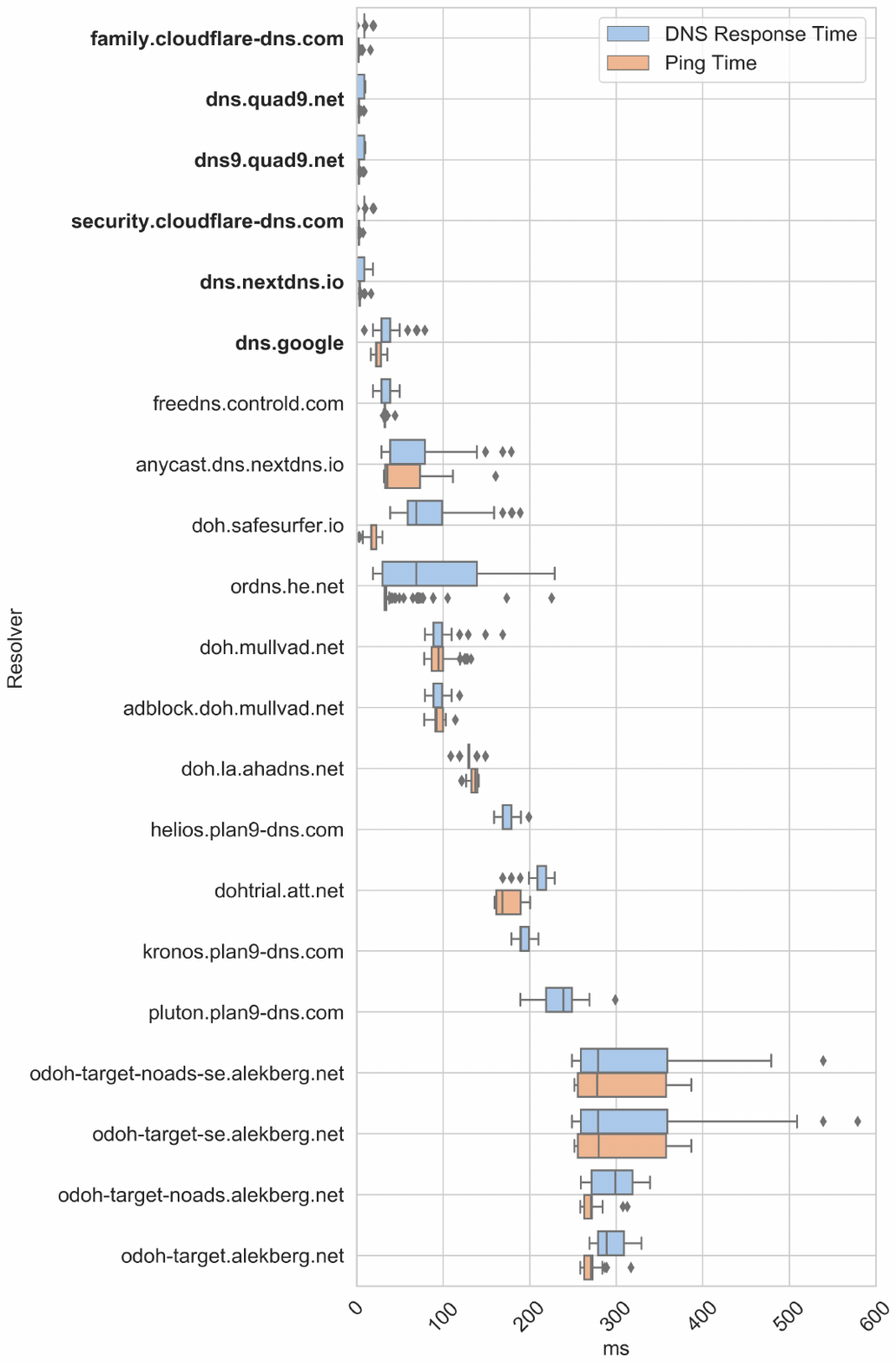}
\caption{Seoul EC2.}
\end{subfigure}
\caption{The DNS response time and ICMP ping time distributions for
    encrypted DNS resolvers located in North America, measured from global vantage points.
    Mainstream resolvers are shown in boldface across all three
    sub-figures.}
\label{fig:dns-NA}
\end{figure*}

\begin{figure*}[h!]
\centering
\begin{subfigure}[b]{0.4\textwidth}
\includegraphics[width=\textwidth]{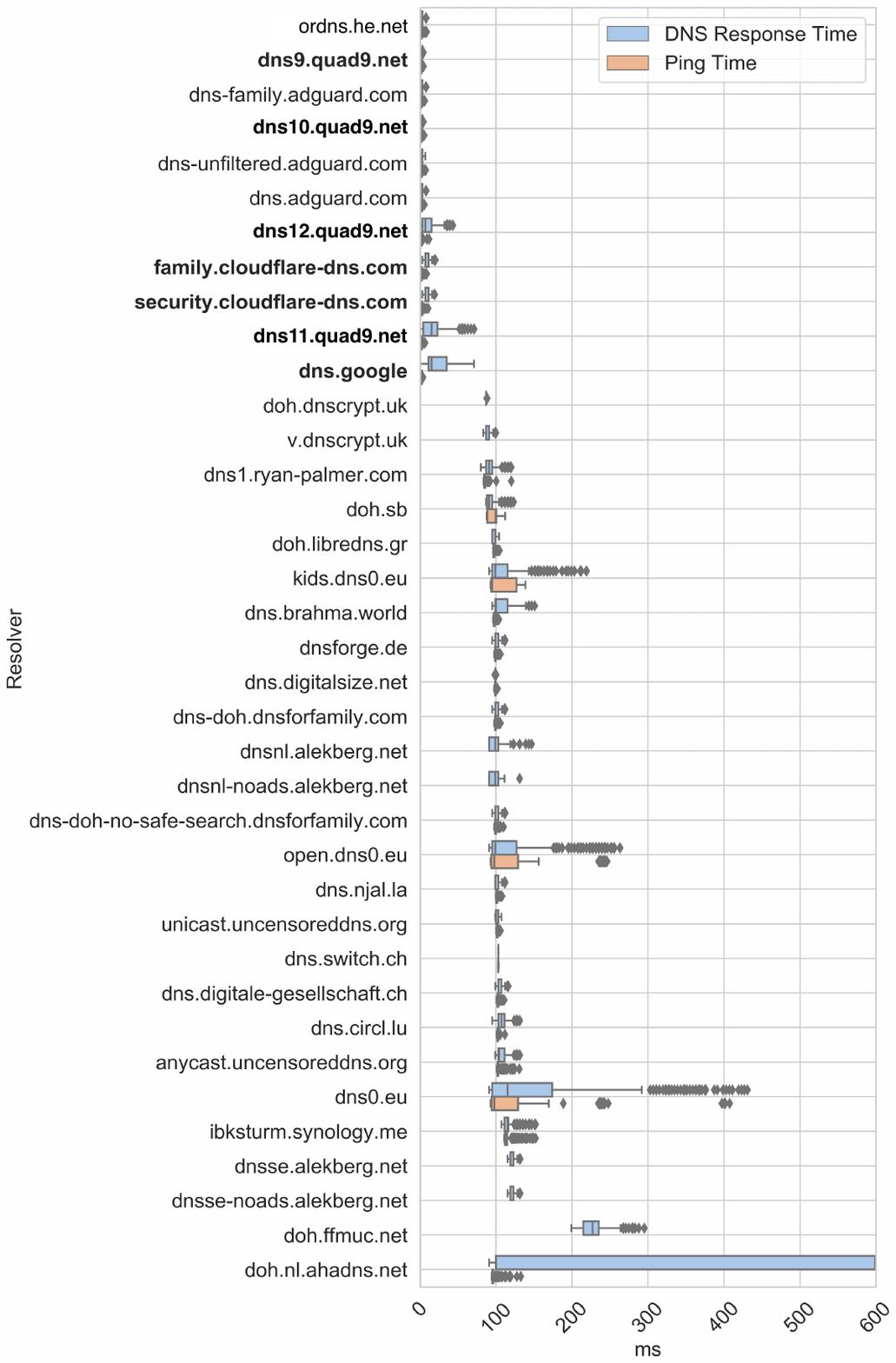}
\caption{U.S. Home Networks.}
\end{subfigure}
\begin{subfigure}[b]{0.4\textwidth}
\includegraphics[width=\textwidth]{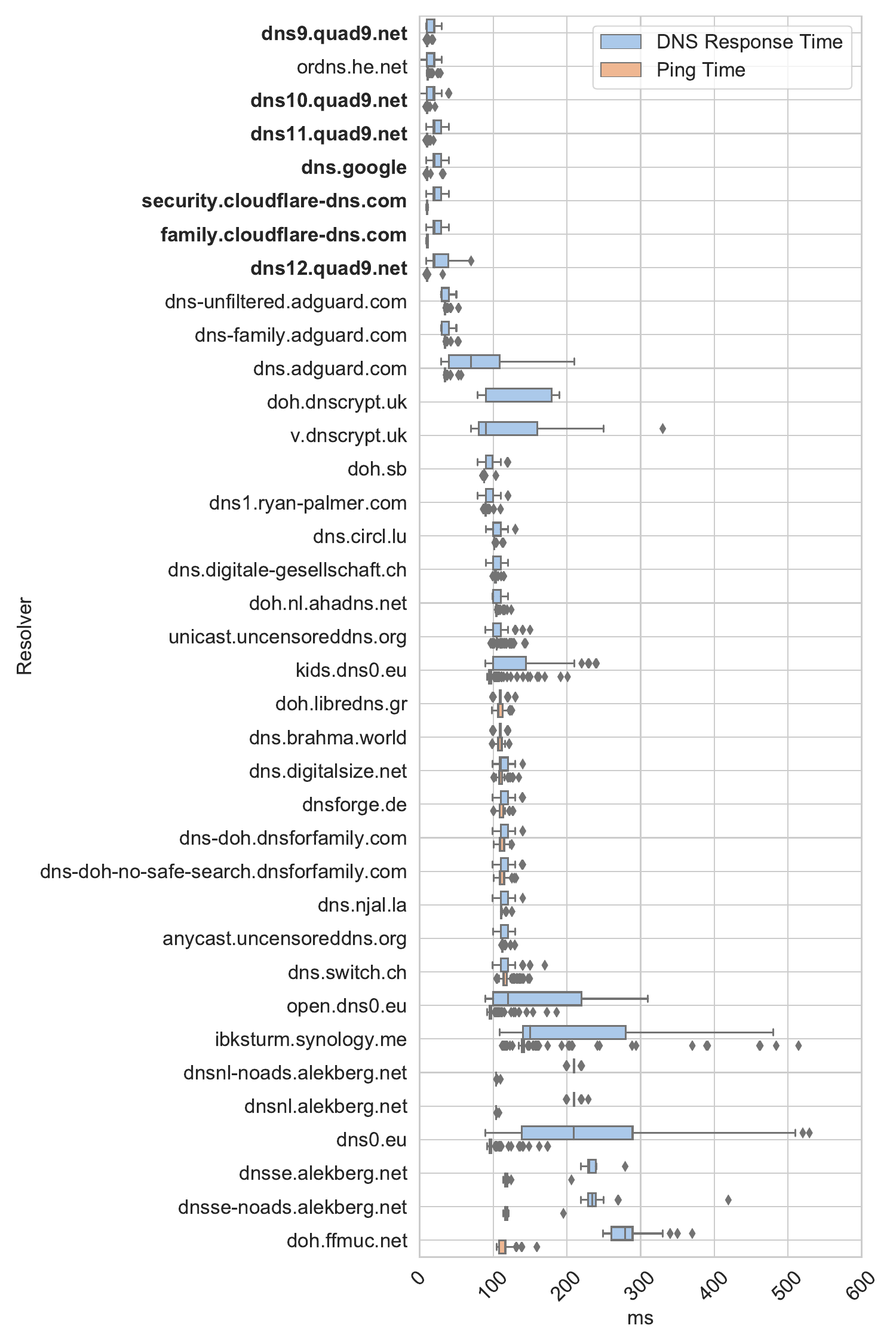}
\caption{Ohio EC2.}
\end{subfigure}
\hfill \\
\begin{subfigure}[b]{0.4\textwidth}
\includegraphics[width=\textwidth]{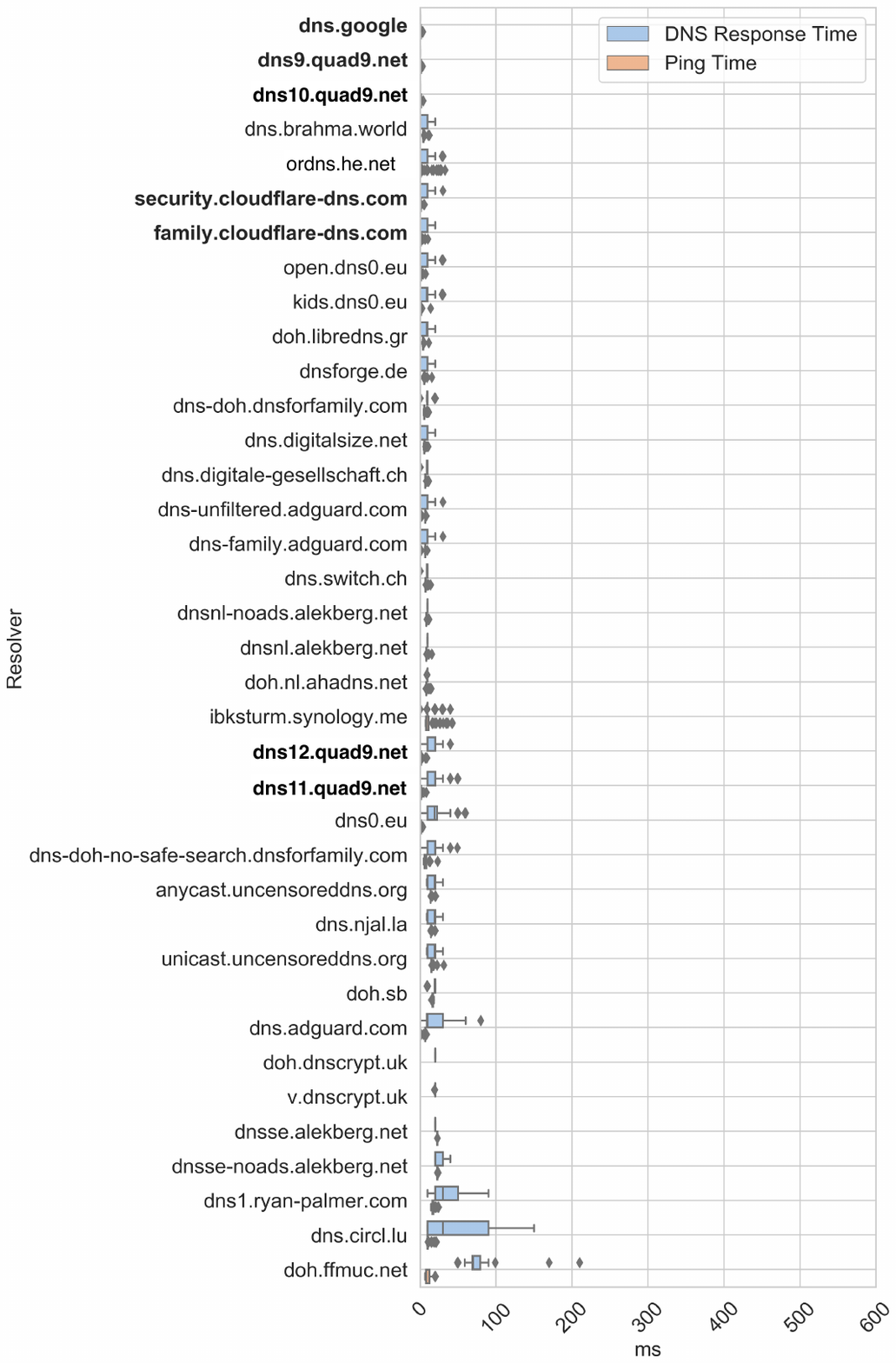}
    \caption{Frankfurt EC2 (Local).}
\end{subfigure}
\begin{subfigure}[b]{0.4\textwidth}
\includegraphics[width=\textwidth]{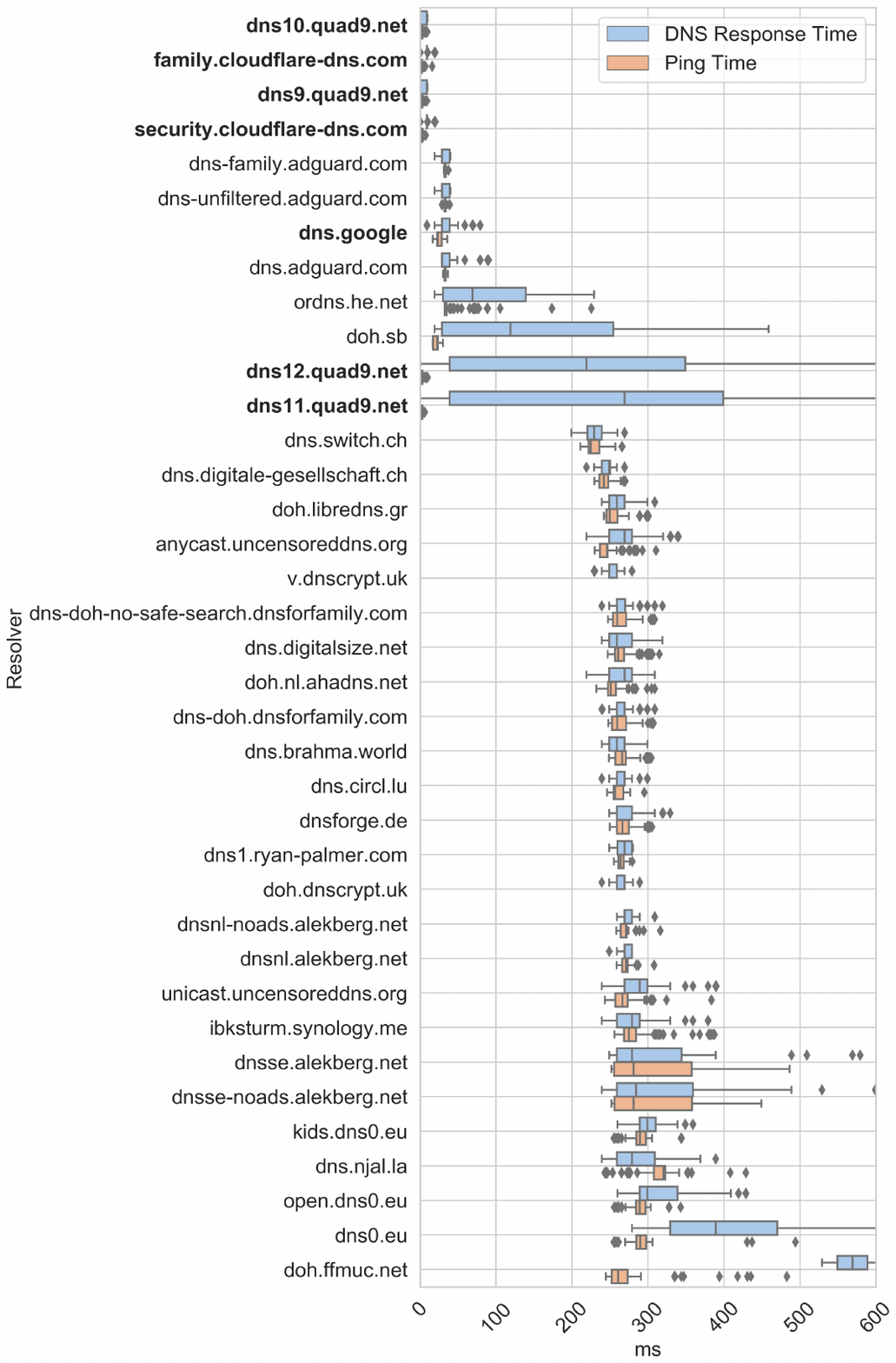}
\caption{Seoul EC2.}
\end{subfigure}
\caption{The DNS response time and ICMP ping time distributions for
    encrypted DNS resolvers located in Europe, measured from global vantage points.
    Mainstream resolvers are shown in boldface across all three
    sub-figures.}
\label{fig:dns-europe}
\end{figure*}
\begin{figure*}[h!]
\centering
\begin{subfigure}[b]{0.4\textwidth}
\includegraphics[width=\textwidth]{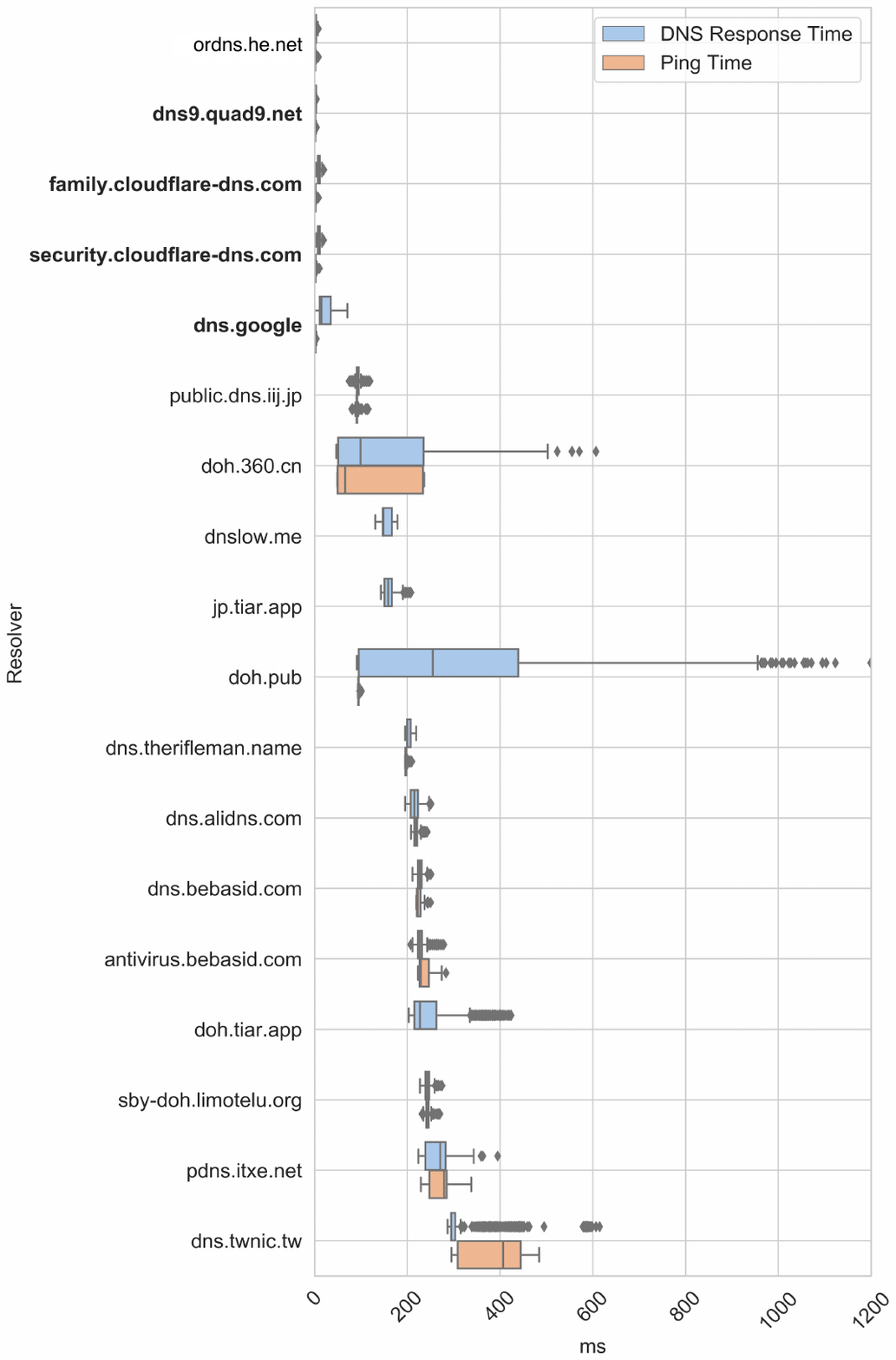}
\caption{U.S. Home Networks}
\end{subfigure}
\begin{subfigure}[b]{0.4\textwidth}
\includegraphics[width=\textwidth]{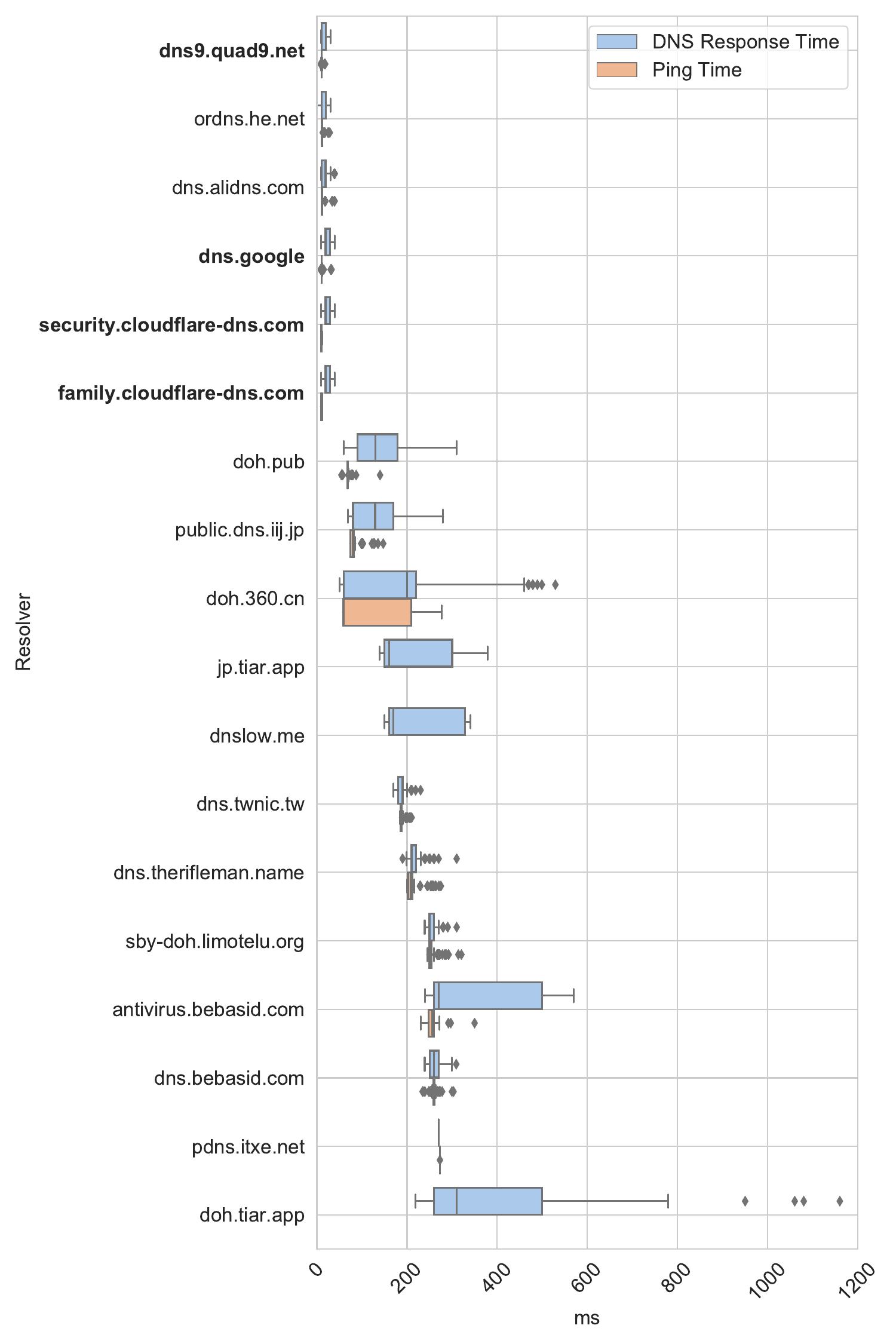}
\caption{Ohio EC2.}
\end{subfigure}
\hfill \\
\begin{subfigure}[b]{0.4\textwidth}
\includegraphics[width=\textwidth]{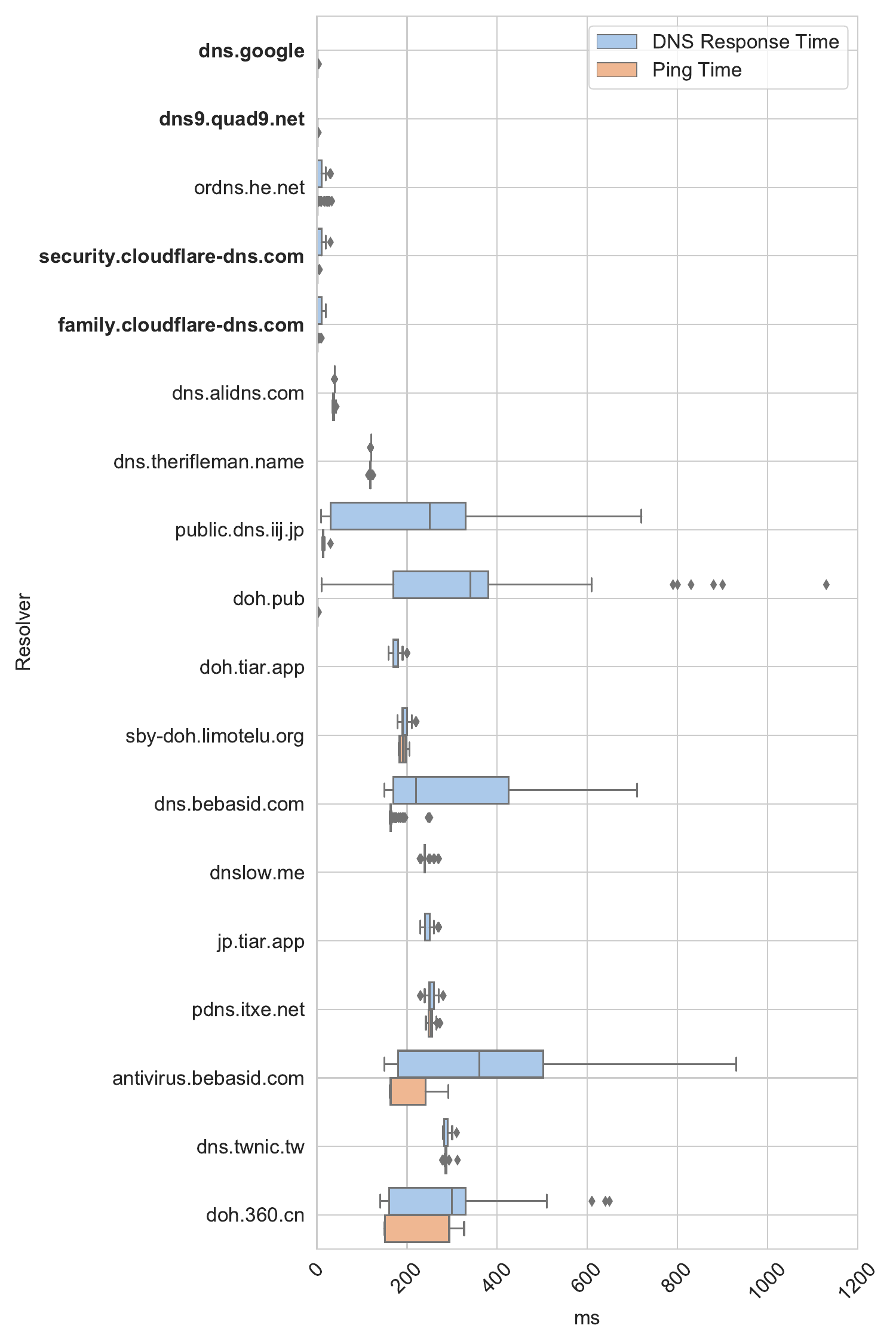}
    \caption{Frankfurt EC2.}
\end{subfigure}
\begin{subfigure}[b]{0.4\textwidth}
\includegraphics[width=\textwidth]{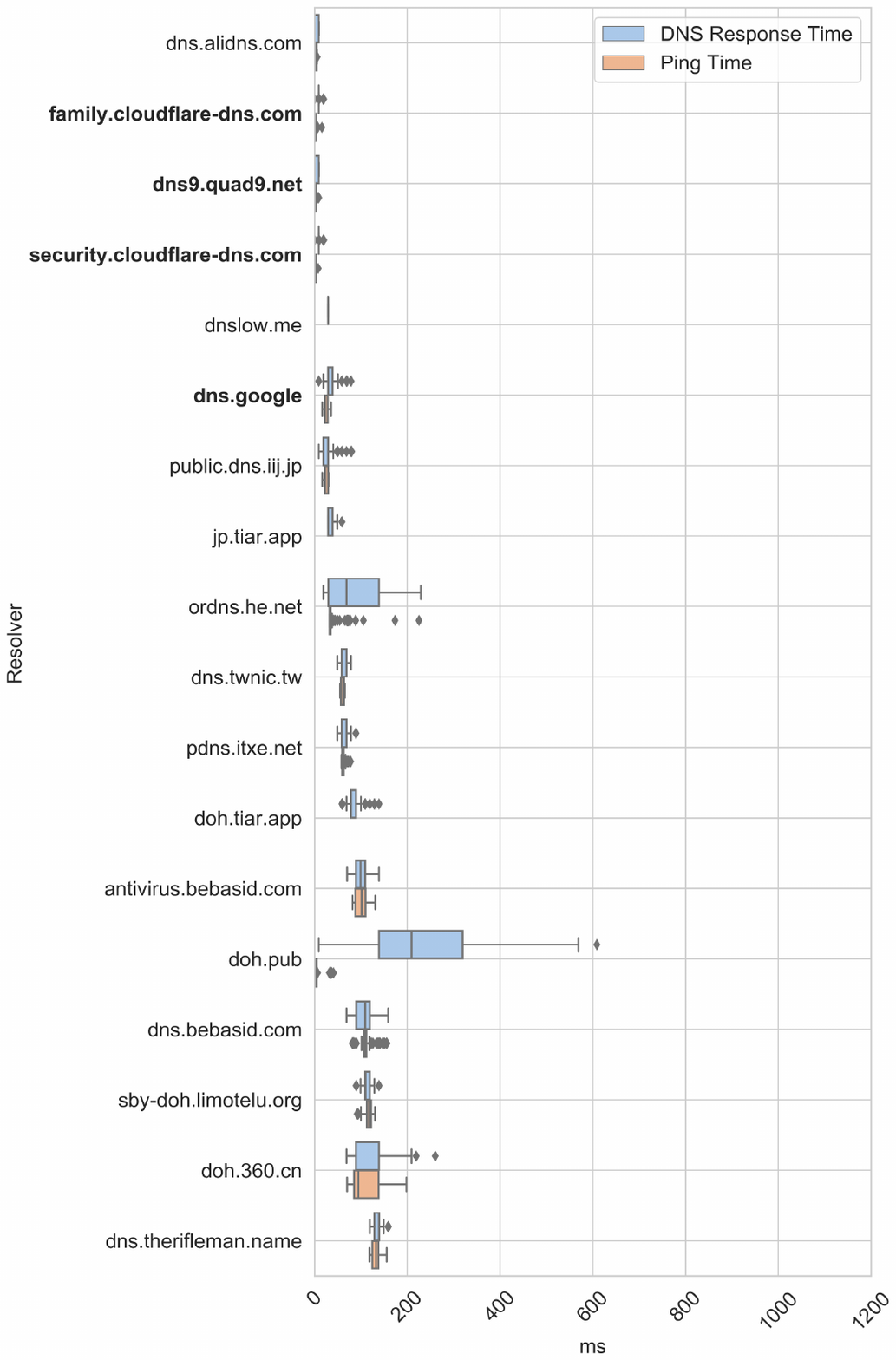}
    \caption{Seoul EC2 (Local).}
\end{subfigure}
\caption{The DNS response time and ICMP ping time distributions for
    encrypted DNS resolvers located in Asia, measured from global vantage points.
    Mainstream resolvers are shown in boldface across all three
    sub-figures.}
\label{fig:dns-asia}
\end{figure*}

\end{sloppypar}
\end{document}